\pdfoutput=0
\documentclass[english,aps,prl,twocolumn]{revtex4}
\usepackage[T1]{fontenc}
\usepackage[latin9]{inputenc}
\usepackage{amstext}
\usepackage{graphicx}

\makeatletter
\@ifundefined{textcolor}{}
{%
 \definecolor{BLACK}{gray}{0}
 \definecolor{WHITE}{gray}{1}
 \definecolor{RED}{rgb}{1,0,0}
 \definecolor{GREEN}{rgb}{0,1,0}
 \definecolor{BLUE}{rgb}{0,0,1}
 \definecolor{CYAN}{cmyk}{1,0,0,0}
 \definecolor{MAGENTA}{cmyk}{0,1,0,0}
 \definecolor{YELLOW}{cmyk}{0,0,1,0}
 }

\makeatother

\usepackage{babel}

\begin{document}

\title{Single-Gate Accumulation-Mode InGaAs Quantum Dot with a Vertically
Integrated Charge Sensor}

\author{E. T. Croke}

\email{croke@hrl.com}

\author{M. G. Borselli }

\author{M. F. Gyure }

\author{S. S. Bui }

\author{I. I. Milosavljevic }

\author{R. S. Ross }

\author{A. E. Schmitz }

\author{A. T. Hunter }

\affiliation{HRL Laboratories LLC, Malibu, CA 90265, USA}

\date{\today}
\begin{abstract}
We report on the fabrication and characterization of a few-electron
quantum dot controlled by a single gate electrode. Our device has
a double-quantum-well design, in which the doping controls the occupancy
of the lower well while the upper well remains empty under the free
surface. A small air-bridged gate contacts the surface, and is positively
biased to draw laterally confined electrons into the upper well. Electrons
tunneling between this accumulation-mode dot and the lower well are
detected using a quantum point contact (QPC), located slightly offset
from the dot gate. The charge state of the dot is measured by monitoring
the differential transconductance of the QPC near pinch-off. Addition
spectra starting with $N=0$ were observed as a function of gate voltage.
DC sensitivity to single electrons was determined to be as high as
$8.6$\%, resulting in a signal-to-noise ratio of $\sim9:1$ with
an equivalent noise bandwidth of $12.1$~kHz. Analysis of random
telegraph signals associated with the zero to one electron transition
allowed a measurement of the lifetimes for the filled and empty states
of the one-electron dot: $0.38$~ms and $0.22$~ms, respectively,
for a device with a $10$~nm AlInAs tunnel barrier between the two
wells.
\end{abstract}
\maketitle
Semiconductor quantum dots have been a subject of recent interest,
due in part to their potential use in quantum information processing
applications \cite{DiVincenzo,Kouwenhowen,Elzerman,Hanson}. The electron
spin in a magnetic field is a promising two-level quantum system in
which to represent the qubit, as decoherence times can be quite long
(tens to hundreds of milliseconds), particularly in a Si-based implementation
\cite{Eriksson,Tyryshkin}.

In this paper, we describe the practical realization of a novel device
geometry, similar in some respects to devices described in Refs. \cite{Vrijen,IEEE,Kosaka,Zaitsu},
in which the quantum dot and the readout channel are integrated in
a vertical structure. In our case, the quantum dot is located \emph{above}
the channel instead of adjacent, as is the case for laterally-depleted
dots \cite{Petta}. The design is particularly simple in that it uses
a single gate electrode to define the dot and control its occupancy.
No etching is necessary to define the dot.

\begin{figure}[!h]
\begin{centering}
\includegraphics[bb=0bp 0bp 303bp 436bp,width=2.7in]{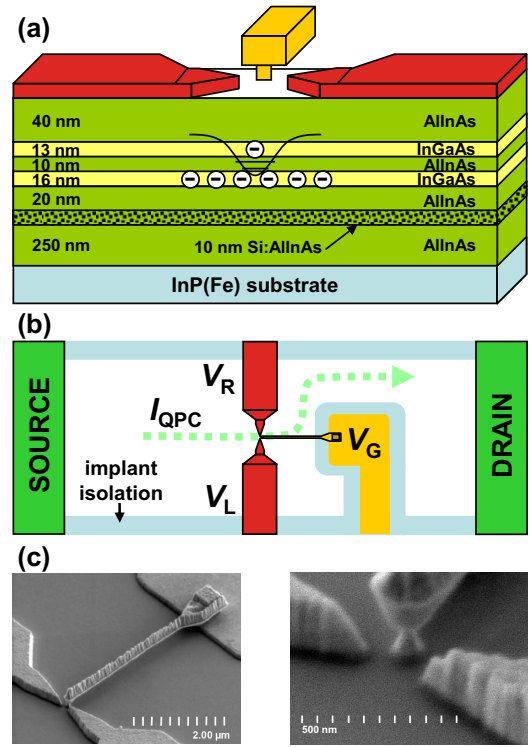}
\par\end{centering}

\caption{\label{fig:Schematic}(a) A schematic diagram showing the layer structure
and gate geometry for our devices. Two In$_{0.53}$Ga$_{0.47}$As
quantum wells surrounded by Al$_{0.48}$In$_{0.52}$As form the confining
layers for both the quantum dot (in the upper well) and the readout
channel (lower well). (b) A view from above showing the air-bridged
gate electrode, the QPC, and the current path. (c)~Scanning electron
microscope images from representative devices.}

\end{figure}

The heteroepitaxial stack, doping profile, and electrode layout were
optimized in a comprehensive theoretical effort described partly in
Ref. \cite{Caflisch}. A schematic diagram detailing the layer structure
and device design is shown in Fig.~\ref{fig:Schematic}(a). The samples
are doped so that away from the gates, under the free surface, the
upper well is fully depleted and the lower well contains a two-dimensional
electron gas (2DEG). Placement of the dopant supply layer below the
two quantum wells was found to result in a more robust design, requiring
less control of the total doping to achieve the desired band profile.
The 2DEG in the lower well also serves as a source of electrons for
the dot.

The quantum dot is located at the center of an implant-isolated region
contacted at each end with sintered ohmic contacts. Lateral depletion
gates under reverse bias form a quantum point contact (QPC) in the
lower well near the dot. The quantum dot is formed in the upper well
by placing the dot gate in forward bias (hence, our use of the term
accumulation mode) to locally push the conduction band in the upper
well below the Fermi level of the 2DEG. Electrons can then tunnel
into the potential well created by the gate when the 2DEG Fermi level
aligns with the quantum states of that localized potential. The typical
device we studied was a nominally $100$~nm diameter dot gate within
$100$~nm of the QPC.

The sample was grown on a $3$~in. semi-insulating, $(100)$-oriented
InP substrate in a gas source MBE system. Doping was adjusted to provide
a $4$~K sheet electron density of $5.2\times10^{11}$\ cm$^{\text{-2}}$;
mobility was measured to be $5.9\times10^{4}$ cm$^{\text{2}}$/volt-sec.
Figure~\ref{fig:Schematic}(b) shows the layout looking down from
above. The air-bridge extended from the gate feed {[}yellow region
surrounded by implant isolation in Fig.~\ref{fig:Schematic}(b){]},
over the semiconductor, contacting it at only one point near the center
of the QPC {[}see also Fig.~\ref{fig:Schematic}(c) for scanning
electron micrograph images of representative devices{]}. The gate
feed was defined using standard optical lithography, whereas the QPC,
the dot gate, and the air-bridge were formed using two \textit{e}-beam
lithography and metallization steps.

\begin{figure}
\begin{centering}
\includegraphics[width=2.9in]{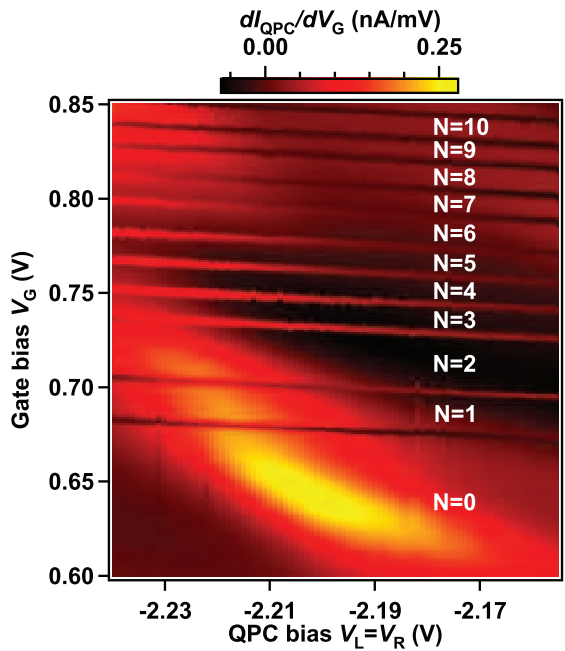}
\par\end{centering}

\caption{\label{fig:Charging}Differential transconductance ($dI_{\textnormal{QPC}}/dV_{\textnormal{G}}$)
obtained by recording the in-phase component of the QPC current due
to a $17.7$~Hz, $1$~mV$_{\textnormal{RMS}}$ excitation added
to $V_{\textnormal{G}}$. Dot transitions are observed as sloping
lines when plotted by $V_{\textnormal{G}}$ versus QPC bias ($V_{\textnormal{L}}=V_{\textnormal{R}}$).
The slight negative slope to the dot transitions is a result of the
capacitive coupling between the dot and the metal QPC gates. Dot occupancy
for the first ten electrons is labeled in white.}

\end{figure}

Devices were introduced into a top-loading $^{3}$He system and tested
at temperatures below $300$~mK. No magnetic fields were used in
the experiments reported here. We measured the differential transconductance
($dI_{\textnormal{QPC}}/dV_{\textnormal{G}}$) of the QPC using a
low-noise transimpedance amplifier (TIA) and standard lock-in techniques.
The differential transconductance signal was generated by applying
a source-drain bias of $500$~$\mu$V across the QPC and adding a
small AC modulation to the DC dot gate voltage, $V_{\textnormal{G}}$.
The lock-in then measured the QPC current at the applied frequency.

Figure~\ref{fig:Charging} shows the amplitude of the in-phase component
of the differential transconductance for a typical device, measured
near QPC pinch-off, as a function of $V_{\textnormal{G}}$ and the
QPC bias $V_{\textnormal{L}}=V_{\textnormal{R}}$ (for the `left'
and `right' QPC gates). For each QPC bias, the gate voltage was swept
from $0.60$ to $0.85$~V, causing electrons to tunnel sequentially
from the lower well into the upper well at particular values of bias
for which the 2DEG Fermi level aligned with the quantum states of
the dot. Transitions are indicated by a dip (peak) when the background
transconductance signal is positive (negative). The dot occupancy
$N$, labeled in white, is evident from the fact that no transitions
are observed below $0.67$~V despite the QPC still being sensitive
to changes in the local electrostatic charge configuration. Furthermore,
the tunneling rate is largely independent of all gate biases used
in the experiment, in contrast to the more common depletion-gated
devices in which pinch-off of the tunnel barriers is always a concern
when the dot is nearly empty \cite{Elzerman,Ciorga}.

\begin{figure}
\begin{centering}
\includegraphics[width=2.7in]{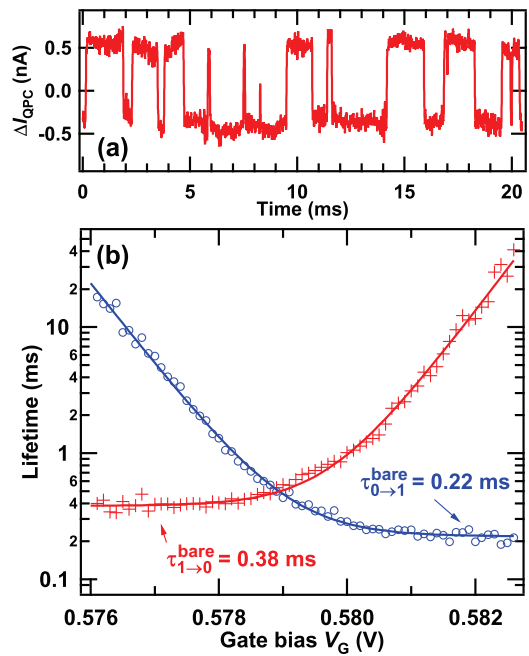}
\par\end{centering}

\caption{\label{fig:RTS}(a) Representative time trace of the change in QPC
current as the dot experiences thermally-induced loading and unloading
events, known as random telegraph signal. The gate bias, $V_{\textnormal{G}}$,
was nominally centered on the $0\leftrightarrow1$ electron transition.
(b) Average lifetimes for loading ($0\rightarrow1$, blue circles)
and unloading ($1\rightarrow0$, red crosses) the first dot state
as a function of $V_{\textnormal{G}}$. The solid lines are fits of
the data to the formulas in the text.}

\end{figure}

From the slope of the transition lines, we obtained the ratio, $C_{\textnormal{G}}/C_{\textnormal{QPC}}\approx10$,
for the gate to QPC capacitances, confirming the dot potential was
primarily controlled by a single gate. The large voltage gap at $N=2$
is a fingerprint of strong electron quantization in the dot potential,
with the scale of orbital energies being comparable to the electrostatic
charging energies. Real-space few-electron simulations of the ground
state filling spectrum were performed using the grown epitaxial structure
and final metal dimensions. Excellent agreement was found between
the simulated and measured voltage gaps providing an estimate for
the charging energy of $\sim5$$\,$meV.

In another experiment, we digitized the output of the TIA as a function
of time for different values of $V_{\textnormal{G}}$. On resonance,
random telegraph signals (RTS) were observed for each transition.
Figure~\ref{fig:RTS}(a) shows a representative time trace of the
QPC current for the $0\leftrightarrow1$ electron transition after
optimizing the QPC bias for maximum sensitivity. The red curve is
the output of the TIA after being low-pass filtered by a double-pole
buffered RC filter set to have a $10$--$90$\% rise time of $30$~$\mu$s.
The measured equivalent noise bandwidth in this configuration was
found to be $12.1$~kHz, resulting in a total of $100$~pA$_{\textnormal{RMS}}$
of noise. Single electron occupation changes resulted in a $900$~pA
change in the $10.5$~nA DC current ($8.6$\% sensitivity, $S/N\approx9$).
Figure~\ref{fig:RTS}(b) is a plot of the average lifetimes (average
of the times spent in the high and low current states, respectively)
for the $0\leftrightarrow1$ electron transition as a function of
$V_{\textnormal{G}}$; the filling voltage is reduced from that of
Fig.~\ref{fig:Charging} because the data were taken after a separate
cooldown. For each gate bias, a histogram of the individual tunneling
times was generated from $5$~seconds of current data, sampled every
$10$~$\mu$s. A simple thresholding procedure was used to determine
when loading and unloading events had occurred. The tunneling times
were confirmed to obey Poisson statistics by both fitting the distribution
of tunneling times to an exponential and observing that the average
lifetime was equal to the standard deviation. The solid lines are
fits to the functions $\tau_{\textnormal{0\ensuremath{\rightarrow}1}}(V_{\textnormal{G}})=\tau_{\textnormal{0\ensuremath{\rightarrow}1}}^{\textnormal{bare}}\left[1+e^{-\alpha(V_{\textnormal{G}}-V_{\textnormal{G}0})/k_{\textnormal{B}}T}\right]$
and $\tau_{\textnormal{\textnormal{1\ensuremath{\rightarrow}0}}}(V_{\textnormal{G}})=\tau_{\textnormal{\textnormal{1\ensuremath{\rightarrow}0}}}^{\textnormal{bare}}\left[1+e^{+\alpha(V_{\textnormal{G}}-V_{\textnormal{G}0})/k_{\textnormal{B}}T}\right]$,
where $\alpha$ is the lever arm of the top-gate to dot potential
\cite{Weis}, $k_{\textnormal{B}}$ is Boltzmann\textquoteright{}s
constant, and $V_{\textnormal{G}0}$ is the gate bias that aligns
the chemical potentials of the empty dot and the lower well. The fitting
parameters, $\tau_{\textnormal{0\ensuremath{\rightarrow}1}}^{\textnormal{bare}}$
and $\tau_{\textnormal{1\ensuremath{\rightarrow}0}}^{\textnormal{bare}}$,
are the minimum lifetimes corresponding to loading and unloading of
the one-electron dot, respectively. The fits yield $k_{\textnormal{B}}T/\alpha=0.67$~mV,
$V_{\textnormal{G}0}=0.5793$~V, and minimum lifetimes of $0.22$~ms
and $0.38$~ms for loading and unloading, respectively. Loading is
approximately a factor of two faster than unloading due to the fact
that both spin states are available to load into when the dot is empty.
These results were obtained for a dot with a 10 nm Al$_{0.48}$In$_{0.52}$As
tunnel barrier between the two wells.

In summary, we have demonstrated the creation of a well-controlled,
few-electron quantum dot in an accumulation mode device using a single
surface gate placed in forward bias. We have also demonstrated single-shot
readout of the charge state of these dots using an adjacent QPC formed
in the lower well as a charge sensor. This device layout offers several
advantages over existing designs. Unlike other vertically-oriented
devices formed by etching pillar structures, it should be straightforward
to place these dots in close proximity and control their coupling
through the use of one additional surface gate placed between the
dot gates. Unlike depletion-gate geometries that require several gates
to form even a single dot, our approach forms the dot and controls
electron occupancy with a single, forward-biased gate.
\begin{acknowledgments}
The authors gratefully acknowledge Profs. HongWen Jiang and Robert
Schwartz for many useful discussions. Sponsored by United States Department
of Defense. The views and conclusions contained in this document are
those of the authors and should not be interpreted as representing
the official policies, either expressly or implied, of the United
States Department of Defense or the U.S. Government. Approved for
public release, distribution unlimited.\end{acknowledgments}

\end{document}